\documentclass[useAMS,usenatbib]{mn2e}
\usepackage[english]{babel}
\usepackage{amssymb}
\usepackage{natbib}
\usepackage{times}
\usepackage{graphics,epsfig}
\usepackage{graphicx}
\usepackage{amsmath}
\usepackage{supertabular}
 
\usepackage{color}

\newcommand{\kms}{km~s$^{-1}$}

\newcommand{\hh}{\ensuremath{\rm H_{2}}}

\newcommand{\mspcc}{\ensuremath{\rm M_{\odot} pc^{-3} }}
\newcommand{\mspc}{\ensuremath{\rm M_{\odot} pc^{-2} }}
\newcommand{\HI}{H{\sc i }}

\begin{document}

\title {Molecular scale height in NGC 7331}
\author [N. N. Patra]{	Narendra Nath Patra$^{1}$ \thanks {E-mail: narendra@ncra.tifr.res.in} \\
	$^{1}$ National Centre for Radio Astrophysics, Tata Institute of Fundamental Research, Pune University campus, Pune 411 007, India\\	
}
\date {}
\maketitle

\begin{abstract}

Combined Poisson's-Boltzman equations of hydrostatic equilibrium were set up and solved numerically for different baryonic components to obtain the molecular scale height as defined by the Half Width at Half Maxima (HWHM) in the spiral galaxy NGC 7331. The scale height of the molecular gas was found to vary between $\sim 100-200$ pc depending on the radius and assumed velocity dispersion. The solutions of the hydrostatic equation and the observed rotation curve were used to produce a dynamical model and consequently a simulated column density map of the molecular disk. The modelled molecular disk found to match with the observed one reasonably well except at the outer disk regions. The molecular disk of NGC 7331 was projected to an inclination of 90$^o$ to estimate its observable edge-on thickness (HWHM), which was found to be $\sim 500$ pc. With an HWHM of $\sim 500$ pc, the observed edge-on extent of the molecular disk was seen to be $\sim$ 1 kpc from the mid-plane. This indicates that in a typical galaxy, hydrostatic equilibrium, in fact, can produce a few kilo-parsec thick observable molecular disk which was thought to be difficult to explain earlier.
\end{abstract}


\section{Introduction}
\label{intro}
                                                                                                                                                                                                                                                                                                                                                                                                                                                                                                                                                                                                                                                                                                                                                                                                                                                                                                                                                                                                                                                                                                                                                                                                                                                                                                                                                                                                                                     Molecular gas plays a significant role in galaxy formation and evolution. Molecular clouds
 provide the sites of active star formation and hence host a suite for                                                                                                                                                                                                                                                                                                                                                                                                                                                                                                                                                                                                                                                                                                                                                                                                                                                                                                                                                                                                                                                                                                                                                                                                                                                                                                                                                                                                                                                                                                                                                                                                                                                                                                                                                                                                                                                                                                                                                                                                                                                                                                                                                                                                                                                                                                                                                                                                                                                                                                                                                                                                                                                                                                                                                                                                                                                                                                                                                                                                                                                                                                                                                                                                                                                                                                                                                                                                                                                                                                                                                                                                                                                                                                                                                                                                                                                                                                                                                                                                                                                                                                                                                                                  stellar activity, e.g. stellar feedback, supernova etc. The conversion of the gas into stars significantly regulated by this phase of the ISM. Hence, the abundance and distribution of the molecular gas in galaxies are of immense importance. Dynamically, the molecular gas in galaxies is expected to settle in a thin disk due to its low thermal pressure; however, its three-dimensional distribution is still poorly understood.

It is challenging to measure the three-dimensional distribution of the molecular clouds in the Galaxy mainly due to distance ambiguities and high opacity of $^{12} CO$ emission. Out of the plane distribution of molecular gas outside the solar circle was first reported by \citet{grabelsky87} using large-scale $^{12} CO$ survey. Extending on their study, \citet{wouterloot90} used IRAS point source catalogue and $^{12} CO$ observations to identify molecular clouds and their distribution in the outer Galaxy. In the inner Galaxy, the extinction is significant, and the distances to the molecular clouds were estimated mainly from their latitude and hence are ambiguous \citep[e.g.,][]{sanders84,scoville87,bronfman88}. From these studies, one can only conclude that most of the molecular gas resides in a thin disk of scale height (HWHM) $\sim$ 75 pc in the inner Galaxy and can flare up to $200-250$ pc at the outer Galaxy.

However, many recent studies provide mounting evidence of a much thicker molecular disk than what was expected earlier. For example, \citet{garcia-burillo92} observed NGC 891 using IRAM 30 m telescope and found a thick molecular disk. They detected molecular gas at $\sim 1 - 1.4 $ kpc above and below the mid-plane. \citet{pety13} also shown that the molecular component in the M51 galaxy does contain a diffuse component which has a typical scale height of $\sim 200$ pc. \citet{combes97b} studied two nearly face-on galaxies NGC 628 and NGC 3938 to find out that the velocity dispersion of the molecular and atomic gas are similar indicating a diffuse component of the molecular gas with high-velocity dispersion. \citet{calduprimo13} studied the stacked spectra of \HI and CO in a comprehensive sample of 12 galaxies using THINGS \citep{walter08} and HERACLE \citep{leroy09a} survey data respectively to find that the velocity dispersion of the molecular gas is almost the same as that of the atomic gas. Later, very recently, \citet{mogotsi16} used the same sample to study individual \HI and CO spectra and found a little lower velocity dispersion for the molecular gas than the atomic gas. Nevertheless, the line width of the molecular gas found to be $\sim 8-10$ \kms, which is much higher than the velocity dispersion expected in a thin disk.

Though these studies indicate a thicker molecular disk in external galaxies, they do not provide any detailed three-dimensional distribution. Direct measurement of the same is not possible (even for an edge-on galaxy) due to line-of-sight integration effect. Theoretically, the vertical distribution of the gas disk is determined by the balance between gravity and pressure under hydrostatic equilibrium. Hence, the distribution of the molecular gas (or any other component) can be obtained by solving combined Poisson's-Boltzman equation of hydrostatic equilibrium at any radius. Not only that, the solutions then can be used to produce observables and compared with real observations to constrain various input parameters (e.g. velocity dispersion, inclination etc.) to the hydrostatic equilibrium equation \citep[see, e.g.,][]{patra14}.

Many previous studies used vertical hydrostatic equilibrium condition to estimate the vertical structure of atomic gas in spiral and dwarf galaxies \citep{olling95,becquaert97,narayan05c,banerjee08,banerjee10,banerjee11,patra14}. These studies neglected the contribution of the molecular gas due to lack of observational inputs. With recent surveys of molecular gas in nearby galaxies (e.g. HERACLE \citep{leroy09a}), it is possible to estimate the vertical scale height of the molecular gas in nearby spiral galaxies. However, detecting molecular gas in dwarf galaxies remains challenging, and no significant detections were made in spite of substantial efforts \citep{cormier14,mcquinn12,leroy06}.

In this paper, I set up the combined Poisson's-Boltzman equation of hydrostatic equilibrium for a galactic disk with different baryonic components, i.e., stars, atomic gas and molecular gas under the external potential of dark matter halo and solve for the galaxy NGC 7331. Many aspects of NGC 7331 (e.g., stellar content, star formation rates etc.) are remarkably similar to that of the Galaxy, and sometimes it is referred to as ``the Milky Way's twin''. However, though, there exist a few structural differences as well between them, e.g., the Galaxy is believed to be a barred spiral whereas NGC 7331 doesn't have a bar. These facts make it even more interesting to study this galaxy. Along with that, this galaxy is a part of THINGS survey \citep{walter08} and HERACLE \citep{leroy09a} survey which makes the necessary data available for this study. Not only that this galaxy has an inclination of $\sim 76^o$ which makes it suitable for the estimation of its rotation curve reliably and at the same time its gas disk produces an observed thickness which is sensitive to the vertical structure of the gas disk. I numerically solve the second order coupled differential equations to obtain the vertical structure of the molecular disk. The hydrostatic equilibrium is a crucial assumption for this study. Any violation of this assumption would lead to a wrong interpretation of the data. Many previous studies revealed the existence of star-burst driven molecular outflow and supershells which are expected to disturb the hydrostatic equilibrium \citep{bolatto13,irwin96}. However, these disturbances are mostly restricted to the central region of $\lesssim$ 1 kpc where the star formation rate is much higher than the outer parts of a galaxy. A central region of 2 kpc was excluded in this study to avoid several complications related to hydrostatic equilibrium, and it is expected that the assumption of the hydrostatic equilibrium would be mostly valid for the rest of the disk of NGC 7331. Even though it is possible that in places within the galaxy this assumption might not hold good, but as I am using an azimuthally averaged quantities to estimate the molecular scale height, the local fluctuations are expected to be smoothed away.


\section{Modelling the galactic disks}
\label{model_disc}
\subsection{Formulation of equation}

\begin{figure*}
\begin{center}
\begin{tabular}{c}
\resizebox{\textwidth}{!}{\includegraphics{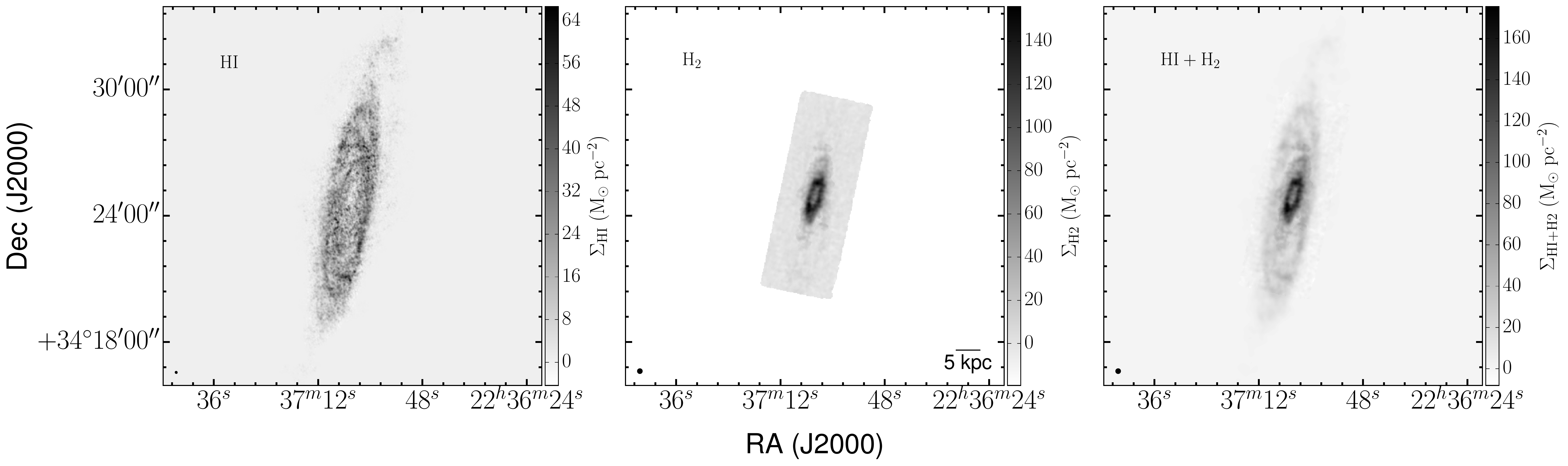}}
\end{tabular}
\end{center}
\caption{The observed column density maps of NGC 7331. Left panel: The \HI column density map from THINGS survey \citep{walter08}. Middle panel: The observed total intensity map of the molecular gas as traced by CO from the HERACLE survey \citep{leroy09a}. Right panel: The intensity distribution of total gas column density, i.e., \HI+$\rm ~H_2$. The colour bars in each panel indicates the observed column densities in the unit of $M_{\odot} pc^{-2}$.}
\label{mom}
\end{figure*}

I model the galactic disk assuming it to be a three component system consisting of stars, atomic gas and molecular gas settled under mutual gravity in the external potential field of the dark matter halo. All the disks of different components would then be in vertical hydrostatic equilibrium individually. For simplicity, all the baryonic disks are considered to be co-planar, concentric and symmetric. Here I set up the hydrostatic equilibrium equation in a cylindrical-polar coordinate (R, z). The observed column density distribution of different baryonic components hence was deprojected (Fig.~\ref{sden}) to obtain the surface density distribution which is in the cylindrical polar coordinate too. The potential of the dark matter halo is considered to be fixed and can be determined observationally (from mass modelling). The Poisson's equation for the disks plus dark matter halo then can be given as

\begin{equation}
\label{eq1}
\frac{1}{R} \frac{\partial }{\partial R} \left( R \frac{\partial \Phi_{total}}{\partial R} \right) + \frac{\partial^2 \Phi_{total}}{\partial z^2} = 4 \pi G \left( \sum_{i=1}^{3} \rho_{i} + \rho_{h} \right)
\end{equation}

\noindent where $\Phi_{total}$ is the total potential due to all the disk components and dark matter halo. $\rho_i$ indicate the volume density of different baryonic components where $\it i$ runs for stars, atomic gas and molecular gas. $\rho_h$ denotes the mass density of dark matter halo. As an NFW profile better describes the dark matter halos of spiral galaxies as compared to an isothermal one \citep{moore98,jing00,deblok08}, I chose to adopt an NFW distribution to represent the dark matter halo of NGC 7331. The dark matter density profile of an NFW halo \citep{navarrofrenkwhite97} can be given as  

\begin{equation}
\label{nfw}
\rho(R) = \frac {\rho_0}{\frac{R}{R_s} \left( 1 + \frac{R}{R_s}\right)^2}
\end{equation}

\noindent where $\rho_0$ is the characteristic density and $R_s$ is the scale radius. These two parameters describe entirely a spherically symmetric NFW dark matter halo. 




The equation of hydrostatic equilibrium for individual components can be written as 

\begin{equation}
\label{eq4}
\frac{\partial }{\partial z} \left(\rho_i {\langle {\sigma}_z^2 \rangle}_i \right) + \rho_i \frac{\partial \Phi_{total}}{\partial z} = 0
\end{equation}

\noindent where ${\langle {\sigma}_z \rangle}_i$ is the vertical velocity dispersion of the $i^{th}$ component, an input parameter. \\

\noindent Eliminating ${\Phi}_{total}$ from Equation (1) and (3),

\begin{equation}
\label{eq5}
\begin{split}
{\langle {\sigma}_z^2 \rangle}_i \frac{\partial}{\partial z} \left( \frac{1}{\rho_i} \frac{\partial \rho_i}{\partial z} \right) &= \\ 
&-4 \pi G \left( \rho_s + \rho_{HI} + \rho_{H_2} + \rho_h \right)\\
&+ \frac{1}{R} \frac{\partial}{\partial R} \left( R \frac{\partial \phi_{total}}{\partial R} \right)
\end{split}
\end{equation}

\noindent where $\rho_s$, $\rho_{HI}$ and $\rho_{H_2}$ are the mass density of stars, \HI and molecular gas respectively. Eq.~\ref{eq5} represents three second-order partial differential equations in the variables ${\rho}_s$, $\rho_{HI}$ and $\rho_{H_2}$.
However the above equation can be further simplified using the fact \citep[see][for more details]{banerjee10}

\begin{equation}
{\left( R \frac{\partial \phi_{total}}{\partial R} \right)}_{R,z} = {(v_{rot}^2)}_{R,z}
\end{equation}

\noindent where ${(v_{rot})}_{R,z}$ is the circular rotation velocity. Assuming a negligible vertical gradient in ${(v_{rot})}_{R,z}$, one can approximate the ${(v_{rot})}_{R,z}$ by the observed rotation curve $v_{rot}$, which is a function of $R$ alone. Thus Eq.~\ref{eq5} reduces to 
 
\begin{equation}
\begin{split}
{\langle {\sigma}_z^2 \rangle}_i \frac{\partial}{\partial z} \left( \frac{1}{\rho_i} \frac{\partial \rho_i}{\partial z} \right) &= \\
&-4 \pi G \left( \rho_s + \rho_{HI} + \rho_{H_2} + \rho_h \right)\\ 
&+ \frac{1}{R} \frac{\partial}{\partial R} \left( v_{rot}^2 \right)
\end{split}
\label{eq_hydro}
\end{equation}

\noindent Eq.~\ref{eq_hydro} represents three coupled, second-order ordinary differential equations in the variables ${\rho}_s$, $\rho_{HI}$ and $\rho_{H_2}$. The solution of Eq.~\ref{eq_hydro} at any radius ($R$) gives the density of these components as a function of $z$. Thus solutions of this equation will provide the three-dimensional density distribution of different disk components.

\subsection{Input parameters}
\label{input_params}

To get the vertical structure of the molecular disk of any galaxy one needs to solve Eq.~\ref{eq_hydro}. In this work, I solve Eq.~\ref{eq_hydro} for the galaxy NGC 7331 to estimate its vertical molecular structure. This galaxy was observed in \HI as part of the THINGS survey \citep{walter08} and the molecular data was taken from the HERACLE survey \citep{leroy09a}. As it is discussed in later sections, an inclination of $\sim$ 76$^o$ of this galaxy favours in comparing the modelled and the observed molecular disk.

In Fig.~\ref{mom}, I show the observed column density images of NGC7331. The left panel shows the column density of \HI gas as observed by the VLA as part of the THINGS survey \citep{walter08}, whereas the middle panel shows the molecular column density map as observed by the 30-meter IRAM telescope as part of the HERACLE survey \citep{leroy09a}. The right panel of Fig.~\ref{mom} shows the total gas column density map i.e., (\HI $\rm + \ H_2$). The black dots show the observing beams at the left bottom corner of the respective panels. The \HI data was at a higher resolution as compared to the \hh~data, hence to get a total gas column density map, I first smoothed the \HI data with a Gaussian kernel to produce an output resolution same as the \hh~map and then sum them together to get the total gas column density map. The grey scale in each panel are in the unit of $M_{\odot} pc^{-2}$. I adopt the same CO(2-1) to $H_2$ conversion factor as given by \citet{leroy08}

\begin{equation}
\Sigma_{H_2} (M_{\odot} pc^{-2}) = 5.5 \ I_{CO}(2 \rightarrow 1) \ (K kms^{-1})
\end{equation}

\begin{figure}
\begin{center}
\begin{tabular}{c}
\resizebox{0.4\textwidth}{!}{\includegraphics{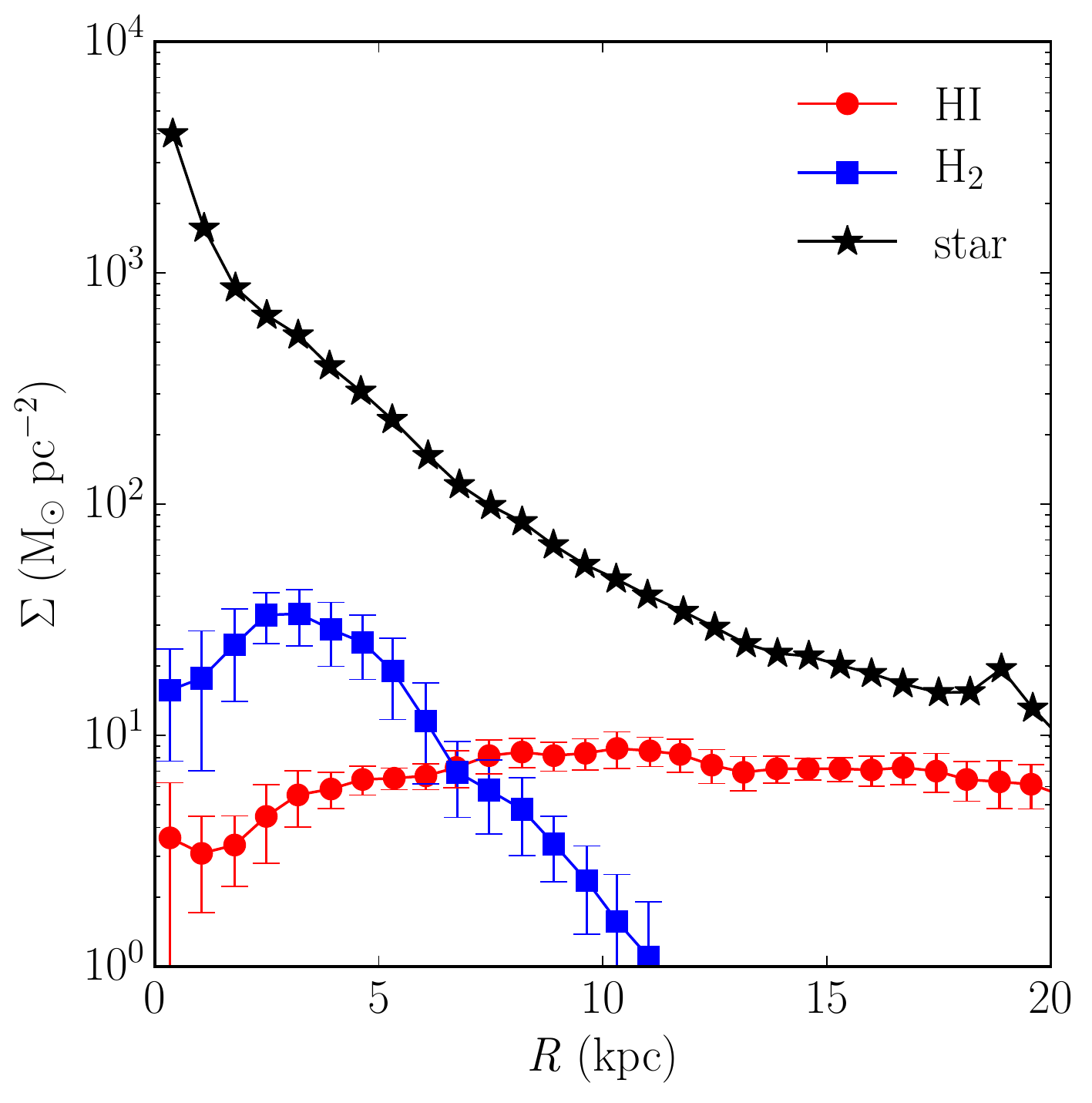}}
\end{tabular}
\end{center}
\caption{The deprojected face-on surface density profiles of NGC7331. The black asterisks represent stellar surface density; the blue squares represent the surface density of molecular gas whereas the red circles represent the \HI surface density. It can be noted that the \hh~disk extends only up to $\sim$ 10 kpc, whereas the \HI and the stellar disks extend up to a much larger radius. The data were taken from \citet{leroy08}.}
\label{sden}
\end{figure}

As we will be solving each component separately at a time (see \S~\ref{solving_tech}), the surface densities of the individual components are of particular interest. In Fig.~\ref{sden}, I plot the deprojected face-on surface densities of different disk components (i.e., stars, \HI and \hh) as a function of radius. These data were taken from \citet{leroy08}. As can be seen from the figure, the molecular gas disk extends up to $\sim$ 10 kpc from the centre whereas the \HI and the stellar disk extends far out to a much larger radius. For details of the surface density calculations, I refer the readers to \citet{leroy08}. It should be noted that the surface densities are one of the primary inputs to the hydrostatic equation and are a proxy to the mass distribution in the vertical direction. 

The vertical velocity dispersions of different disk components are another vital input to Eq.~\ref{eq_hydro}. \citet{banerjee11} shown that the vertical structure of the gaseous components is marginally affected by the accuracy of the assumed stellar velocity dispersion ($\sigma_{s}$). Hence, the stellar velocity dispersion was calculated analytically by assuming an isothermal disk using the formula $\sigma_s \simeq 1.879 \sqrt{l_s \Sigma_s}$ \citep[see][for more details]{leroy08}, where $\sigma_s$ is the stellar velocity dispersion in \kms, $l_s$ is the exponential scale length of the stellar disk in kpc and $\Sigma_s$ is the stellar surface density in \mspc.

The \HI velocity dispersion ($\sigma_{HI}$) in spiral galaxies were studied extensively through \HI spectral line observations. Early work by many authors suggest an \HI velocity dispersion of 6-13 \kms~\citep{shostak84,vanderkruit84,kamphuis93} in galaxies. \citet{petric07} studied the nearly face-on galaxy NGC 1058 to find that the $\sigma_{HI}$ varies between 4-14 \kms~and decreases with radius. In an extensive analysis, \citet{tamburro09} studied the $\sigma_{HI}$ in spiral galaxies from THINGS survey and found a mean \HI velocity dispersion to be $\sim$ 10 \kms at $r_{25}$. In a later study, \citet{ianjamasimanana12}  applied spectral stacking method to the same data to estimate the $\sigma_{HI}$ with higher confidence. They found a $\sigma_{HI} = 12.5 \pm 3.5$ \kms~($\sigma_{HI} = 10.9 \pm 2.1$ \kms~for galaxies with inclination less than 60$^o$). 


\citet{calduprimo13} studied the $\sigma_{HI}$ and $\sigma_{CO}$ using stacking technique in a sample of 12 nearby spiral galaxies. They found $\sigma_{HI}/\sigma_{CO} = 1.0 \pm 0.2$ with a median $\sigma_{HI} = 11.9 \pm 3.1$ \kms. However, \citet{mogotsi16} studied the same sample by analysing individual high SNR spectra to find a $\sigma_{HI}/\sigma_{CO} = 1.4 \pm 0.2$ with $\sigma_{HI} = 11.7 \pm 2.3$ \kms and $\sigma_{CO} = 7.3 \pm 1.7$ \kms. As can be seen from these studies, the $\sigma_{HI}$ in spiral galaxies could be assumed to be $\sim$ 12 \kms. It can be noted that due to their mass budget, the dark matter halo and the stars dominantly decide the gravitational field which is followed by the gas components. In such a situation, the distribution of \HI gas can only marginally influence the distribution of the molecular gas, and hence, the velocity dispersion of \HI has minimal effect on the scale height of the molecular gas.

But, the velocity dispersion of the molecular gas ($\sigma_{H_2}$) directly influences the vertical structure of the molecular disk. \citet{stark84} observed molecular clouds in the Galaxy and found that the velocity dispersion of low mass clouds is higher than the high mass clouds. The low mass clouds have a $\sigma_{H_2} \sim $ 9.0 \kms~whereas the high mass clouds have $\sigma_{H_2} \sim$ 6.6 \kms. \citet{combes97} studied two nearly face-on galaxies to find $\sigma_{H_2} \sim 6-8.5$ \kms. They also found that the $\sigma_{H_2}$ is almost constant over the whole galaxy and comparable to the velocity dispersion of \HI ($\sigma_{HI}$). \citet{stark05} used observations of $^{13}$CO $J = 1 \rightarrow 0$ in 1400 molecular clouds in the Galaxy to find that the velocity dispersion of small clouds is higher than that of the Giant Molecular Clouds (GMCs). I assume a primary velocity dispersion of molecular gas, $\sigma_{H_2}$ to be $\sim$ 7 \kms, along with a variation between 6-10 \kms~to explore the observed molecular disk in more details.

\begin{figure}
\begin{center}
\begin{tabular}{c}
\resizebox{0.4\textwidth}{!}{\includegraphics{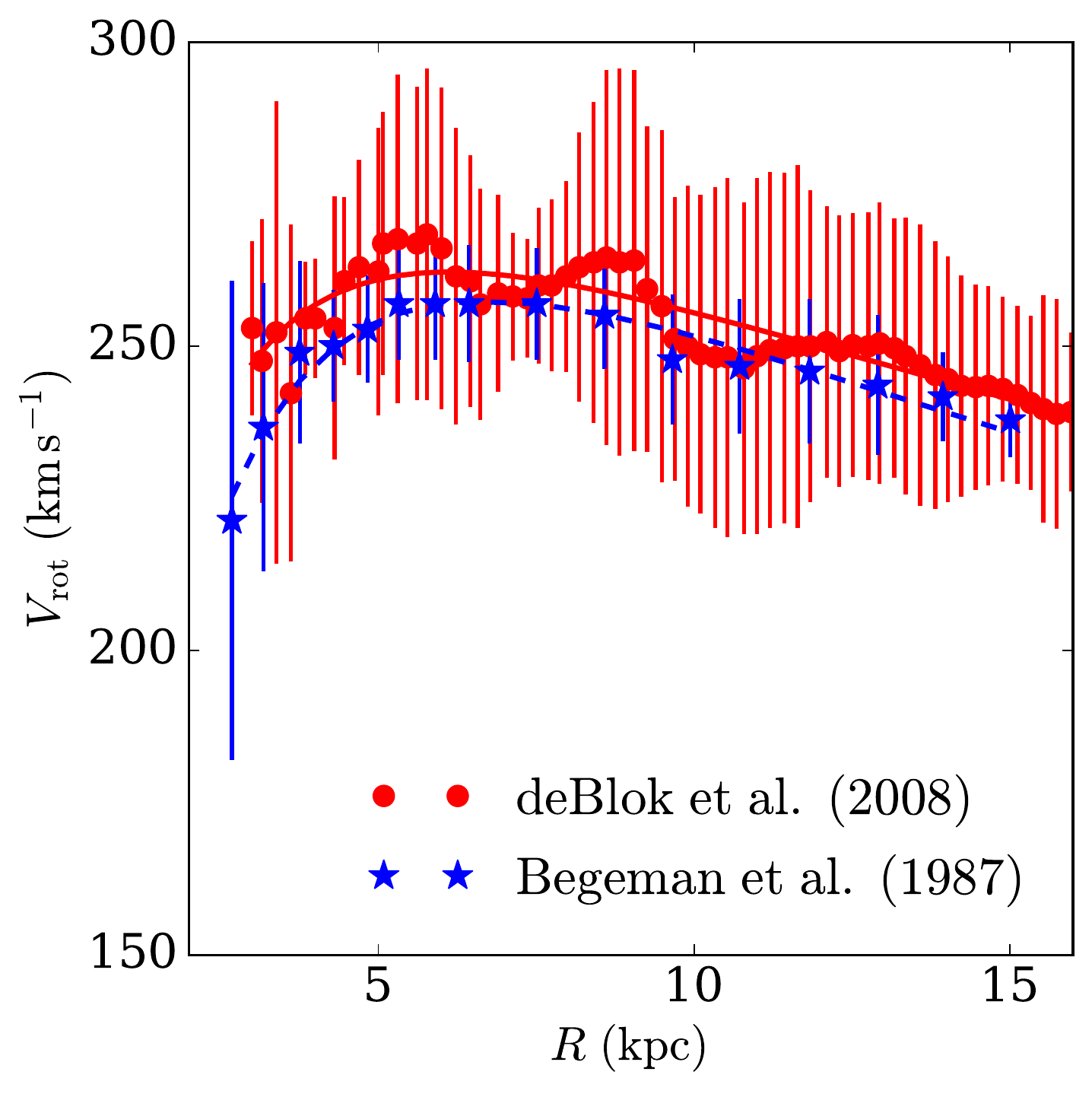}}
\end{tabular}
\end{center}
\caption{The rotation curve of NGC 7331 as measured from \HI data. The solid red circles represent derived rotation curve from THINGS data \citep{deblok08}, whereas the blue asterisks represent data from \citet{begeman87}.}
\label{rotcur}
\end{figure}

The second term on the right-hand side of Eq.~\ref{eq_hydro} represents the contribution of the centripetal acceleration against the gravity. However, the $v_{rot}$ in  Eq.~\ref{eq_hydro} is an observable quantity. In Fig.~\ref{rotcur}, the observed rotation curve of NGC 7331 from \citet{begeman87} and \citet{deblok08} are plotted. \citet{deblok08} used relatively high-resolution \HI data from THINGS survey. As in Eq.~\ref{eq_hydro}, we need to use the derivatives of the rotation velocity ($v_{rot}$), it is useful to parametrise the rotation curve instead of using actual data. A commonly used Brandt's profile \citep{brandt60} given as 

\begin{equation}
v_{rot} (R) = \frac{V_{max}\left(R/R_{max} \right)}{\left(1/3 + 2/3 \left(\frac{R}{R_{max}}\right)^n\right)^{3/2n}}
\end{equation}

\noindent is used to parametrise the rotation curve and the data were fitted to estimate the parameters. A $V_{max} = 262.2 \pm 0.8$ \kms, $R_{max} = 6.1 \pm 0.1$ kpc and $n=0.67 \pm 0.06$ were found for \citet{deblok08} data and $V_{max} = 257.5 \pm 1.0$ \kms, $R_{max} = 6.7 \pm 0.1$ kpc and $n=0.89 \pm 0.07$ were found for \citet{begeman87} data. As can be seen that the fit parameters of both the data matches very well with each other. I chose to work with the parameters found with \citet{begeman87} data as it is smoother than the THINGS data, however, I note that this does not make any fundamental difference to the results.

The dark matter halo is another important input to the hydrostatic equilibrium equation. For NGC 7331 I used the dark matter halo parameters from \citet{deblok08} (Table-4 in their paper). The dark matter halo of NGC 7331 can be described both by the isothermal and NFW profile well. However, as the NFW profile in general describe the dark matter halo of spiral galaxies better than isothermal one \citep{moore98,jing00,deblok08}, I choose to use an NFW profile as given by Eq.~\ref{nfw} with $\rho_0 = 1.05 \times 10^{-3}$ \mspcc~and $r_s = 60.2$ kpc (see \citet{deblok08} for more details).

\subsection{Solving the hydrostatic equilibrium equation}
\label{solving_tech}

With all the inputs mentioned above, Eq.~\ref{eq_hydro} was solved to obtain the vertical structure of different disk components. The coupled second-order ordinary differential equations were solved numerically using 8$^{th}$ order Runge Kutta method as implemented in {\tt scipy} package. As each equation (for individual components) is a second order differential equation, two initial conditions are required to solve it.

\begin{equation}
\left( \rho_i \right)_{z = 0} = \rho_{i,0} \ \ \ \ {\rm and} \ \ \ \frac{d \rho_i}{dz} = 0
\label{init_cond}
\end{equation}

The second boundary condition in the above equation comes from the fact that at the mid-plane, the force, $-\partial \phi_{total}/\partial z$, must be zero due to symmetry in the vertical direction \citep{spitzer42}. Whereas the first boundary condition demands the mid-plane density, $\rho_{i,0}$ to be known a-priory. Though the $\rho_{i,0}$ is not a directly measurable quantity, its value can be estimated using the observed surface density as $\Sigma_i = \int \rho_i(z)  \ dz$. For an individual component, e.g., stars, I first assume a trial $\rho_{s,0}$ and solve Eq.~\ref{eq_hydro} assuming $\rho_{HI} = \rho_{H_2} = 0$. The solution, $\rho_s(z)$ then integrated to obtain the stellar surface density $\Sigma_s$ and compared to the observed one ($\Sigma_s^\prime$) to update the next trial $\rho_{s,0}$. This way, iteratively, an appropriate $\rho_{s,0}$ is estimated such that it produces the observed $\Sigma_s^\prime$ with better than 0.1\% accuracy. It can be noted that using this approach, the surface densities found to converge in less than few hundred iterations. 

However, as in Eq.~\ref{eq_hydro} $\rho_s$, $\rho_{HI}$ and $\rho_{H_2}$ jointly contribute to the gravity (first term in RHS), Eq.~\ref{eq_hydro} have to be solved simultaneously for all the components. We numerically solve it using an iterative approach. We adopt a similar strategy as described in \citet{banerjee08,patra14} (See also \citet{bahcall84b,bahcall84a} for an in-depth analysis of self-gravitating galactic disk systems). In the first iteration, all three equations (for stars, \HI and \hh) are solved independently assuming no coupling (single-component disk). Then these single component solutions are introduced in Eq.~\ref{eq_hydro} to account for the coupling. In every iteration, these solutions are updated till they converge with acceptable accuracy. For example, in the first iteration, Eq.~\ref{eq_hydro} is solved assuming a single-component system and $\rho_s(z)$, $\rho_{HI}(z)$ and $\rho_{H_2}(z)$ is obtained. In the next iteration, the solutions for \HI and \hh, i.e., $\rho_{HI}(z)$ and $\rho_{H_2}(z)$ as obtained in the previous iteration are frozen while solving for stars. Next, solutions for stars, $\rho_{s}(z)$ (updated) and \hh, $\rho_{H_2}(z)$ are frozen while solving for \HI and finally, solutions for stars, $\rho_{s}(z)$ (updated) and \HI, $\rho_{HI}(z)$ (updated) are frozen while solving for \hh. This marks the end of the second iteration. At the end of this iteration, coupled solutions are obtained which is better than the single-disk ones. This process is repeated until the solutions converge with better than 0.1\% accuracy. Physically, in the first iteration, I solve for an individual component, assuming no other components present. Then, in next iterations, any component was solved in the presence of the frozen distribution of other two components (as calculated in the previous iteration). Using this approach, the self-consistent solutions were obtained iteratively starting from the single-component solutions. It can be noted that, for NGC 7331, the solutions were converged in less than ten iterations at any radius. It takes about a few minutes to solve the coupled equation at any radius in a normal workstation. However, as the hydrostatic condition at any radius is independent of other radii, Eq.~\ref{eq_hydro} can be solved in parallel and hence, for fast computation, the hydrostatic equation solver was implemented using MPI based parallel code.

As can be seen from Fig.~\ref{sden}, the surface density of the molecular gas do not extend beyond $R \sim$ 10 kpc and hence Eq.~\ref{eq_hydro} was solved at $R \leq 10$ kpc. From Fig.~\ref{sden}, it can also be noted that the stellar surface density is very high at the central region which indicates a higher energy input to the molecular disk. This might lead to a violation of the hydrostatic assumption in the central region. Not only that but also the assumed dark matter halo density profile (NFW) sharply peaks in the central region. To avoid any divergence due to these factors and a possible non-satisfaction of hydrostatic equilibrium, a central region of 1 kpc was avoided and was not solved. Thus, Eq.~\ref{eq_hydro} was solved within $1 \leq R \leq 10$ kpc with an interval of 100 pc. The linear spatial resolution of the molecular data is $\sim$ 1 kpc. Hence, a radial interval of 100 pc is expected to be more than enough to sample the molecular disk well in the radial direction. However, it is well known that the vertical thickness of the molecular disk is much smaller as compared to its radial extent, and hence a much higher resolution is needed to sample the molecular disk in the vertical direction. To achieve this, an adaptive resolution depending on the scale-height was used and found to be always better than a few pc. It should be emphasised here that a fine resolution in the vertical direction is critical as the molecular disk is very thin. As the molecular gas density vary reasonably in parsec-scale, a grid resolution of few parsecs is necessary to sample the distribution of the molecular gas well in the vertical direction. However, this finely gridded molecular map is convolved with the observed beam to match the observed resolution for comparison.

\section{Results and discussion}
\label{results}

\begin{figure}
\begin{center}
\begin{tabular}{c}
\resizebox{0.4\textwidth}{!}{\includegraphics{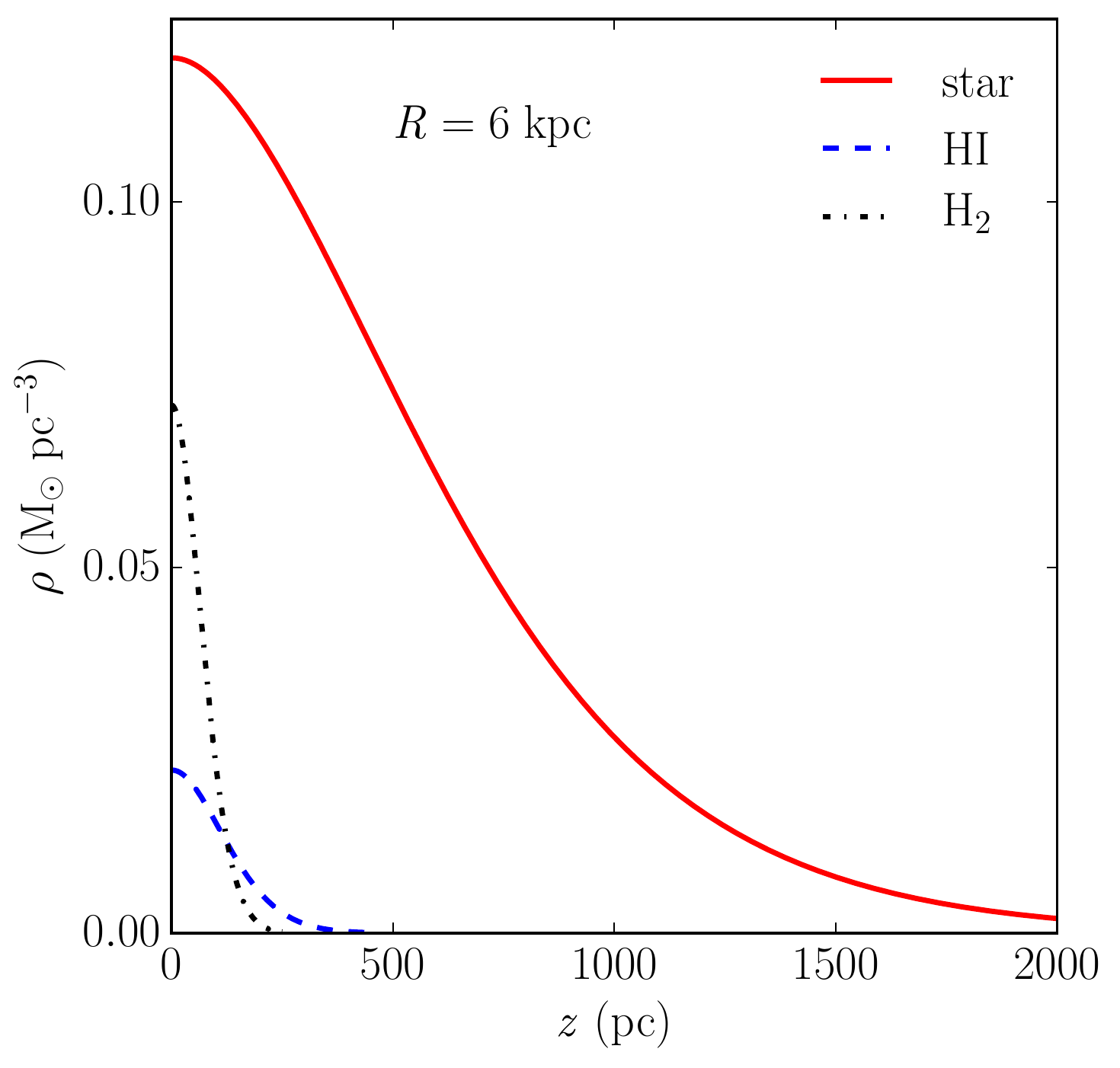}} 
\end{tabular}
\end{center}
\caption{A sample solution of the hydrostatic equilibrium equation. The solid red line represents the stellar mass density as a function of the height from the midplane whereas the blue dashed and black dashed-dotted lines represent the same for \HI and \hh~respectively.}
\label{samp_sol}
\end{figure}

In Fig.~\ref{samp_sol}, sample solutions of Eq.~\ref{eq_hydro}, i.e., the mass density of different disk components (at $R=6$ kpc) are plotted as a function of height (z) from the midplane. It can be seen from the figure, the molecular gas at $R=6$ kpc extends to a much smaller height as compared to the \HI or stellar disk. It should be noted that, for an isothermal single-component disk, the density distribution follows a $sech^2$ law \citep{spitzer42,bahcall84b,bahcall84a}. But due to the coupling of multiple disk components, the solutions deviate from $sech^2$ distribution. In this case, it was found that a Gaussian function can reasonably represent the solutions. The deviation from a $sech^2$ function was seen to be lowest for the stellar disk as it is the gravitationally most dominant component.

The solutions shown in the figure was obtained by solving Eq.~\ref{eq_hydro} assuming a velocity dispersion of 7 \kms~for molecular gas. However, as discussed earlier, $\sigma_{H_2}$ found to vary from galaxy to galaxy or even within a galaxy. To examine how this variation can affect the vertical structure of the molecular disk, Eq.~\ref{eq_hydro} was solved with $\sigma_{H_2}$ varying between 6-10 \kms~in a step of 1 \kms.

\begin{figure}
\begin{center}
\begin{tabular}{c}
\resizebox{0.4\textwidth}{!}{\includegraphics{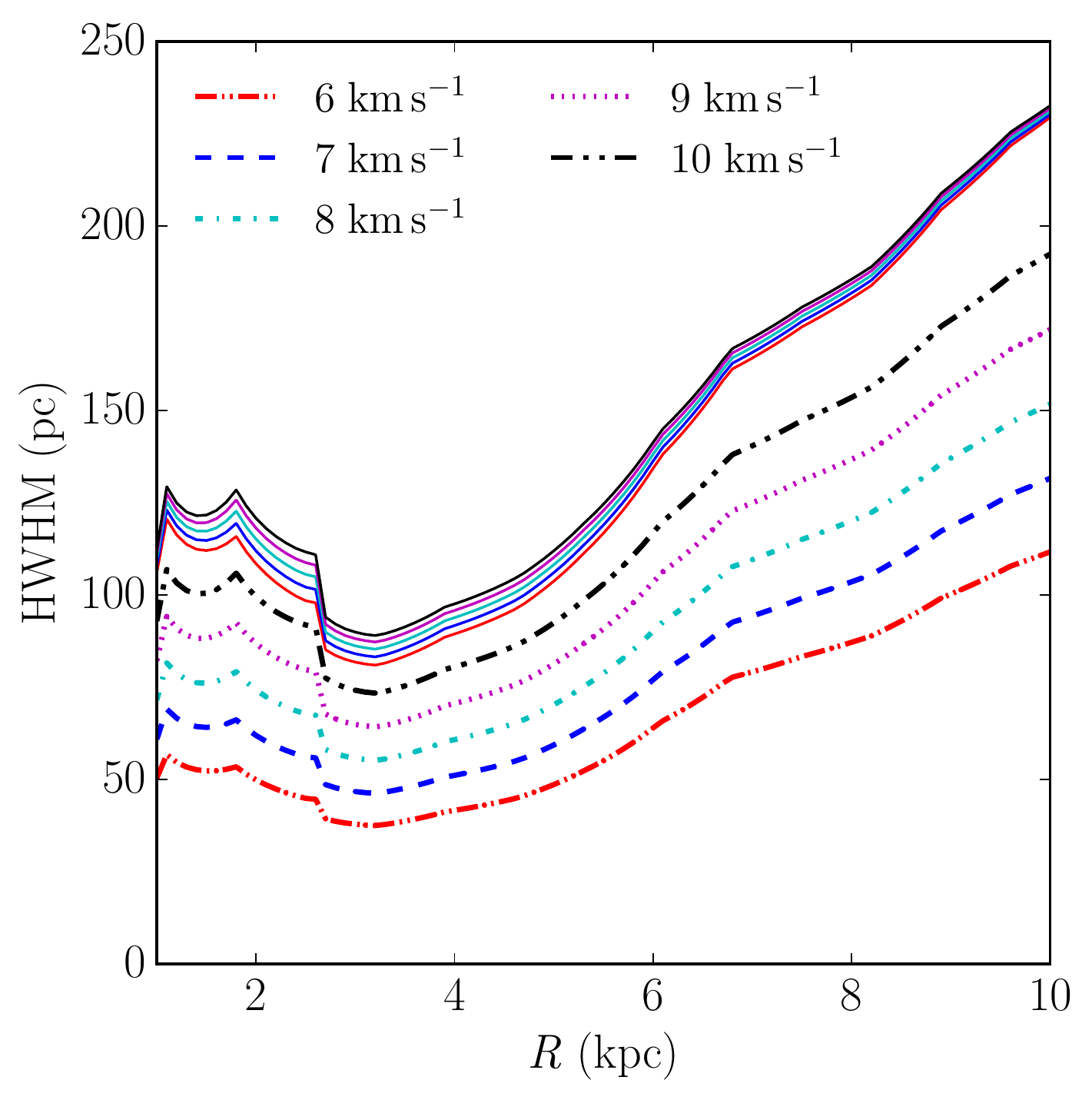}}
\end{tabular}
\end{center}
\caption{The HWHM profiles of the molecular and atomic disks of NGC 7331. The solid lines represent HWHM profiles of the \HI disks, whereas the broken lines represent the same for the molecular disk. Different types of lines represent the scale heights for different assumed $\sigma_{H_2}$ values as quoted in the legend.}
\label{hwhm_3comp}
\end{figure}

A Half width at Half Maxima (HWHM) of the vertical mass density profile was used as a measure of the vertical width of the molecular disk. In Fig.~\ref{hwhm_3comp}, I plot the HWHM profiles of the molecular disk as a function of radius. For comparison, the HWHM profiles of the atomic gas are also plotted in the figure. It can be seen that the molecular scale height in NGC 7331 varies between 50 pc $-$ 200 pc depending on the assumed $\sigma_{H_2}$ and radius. For an assumed $\sigma_{H_2}=$ 7 \kms, the scale height varies between $\sim$ 60 $-$ 100 pc which is $\sim$ a factor of 2 smaller than what is observed in the Galaxy. It can also be seen that the molecular scale height changes by a factor of $\sim$ 2 as one changes the $\sigma_{H_2}$ from 6 \kms~to 10 \kms. Also, the scale height of both the \HI and \hh~increases at $R \lesssim 3$ kpc. This increase is because of strong centripetal acceleration due to the rising rotation curve at these radii (see Fig.~\ref{rotcur}). However, this effect is minimal at the outer radii as the rotation curve significantly flattens at $R \gtrsim 3$ kpc.

The HWHM profile of the molecular disk is not a directly observable quantity even for an edge-on galaxy. Instead, the total intensity map is what is obtained through observation. To check the validity of the derived density distribution of the molecular gas, a three dimensional dynamical model of the molecular disk was produced using the solutions of Eq.~\ref{eq_hydro} and the rotation curve. This 3D model then inclined to the observed inclination of 76$^o$ and convolved with the telescope beam to produce a simulated column density map. This map then can be compared with the observed one to check the consistency of the derived molecular gas density distribution.

\begin{figure}
\begin{center}
\begin{tabular}{c}
\resizebox{0.4\textwidth}{!}{\includegraphics{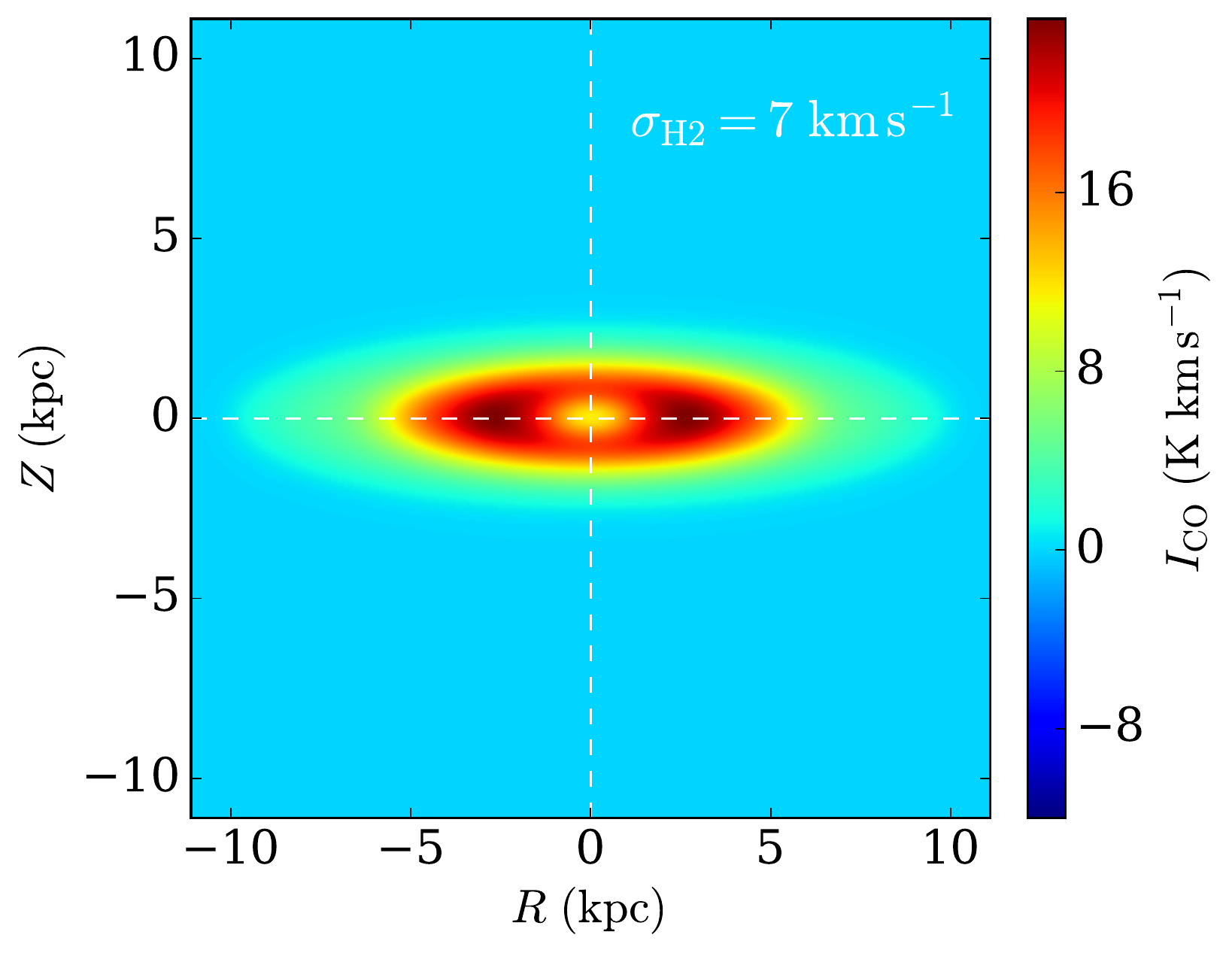}}\\
\resizebox{0.4\textwidth}{!}{\includegraphics{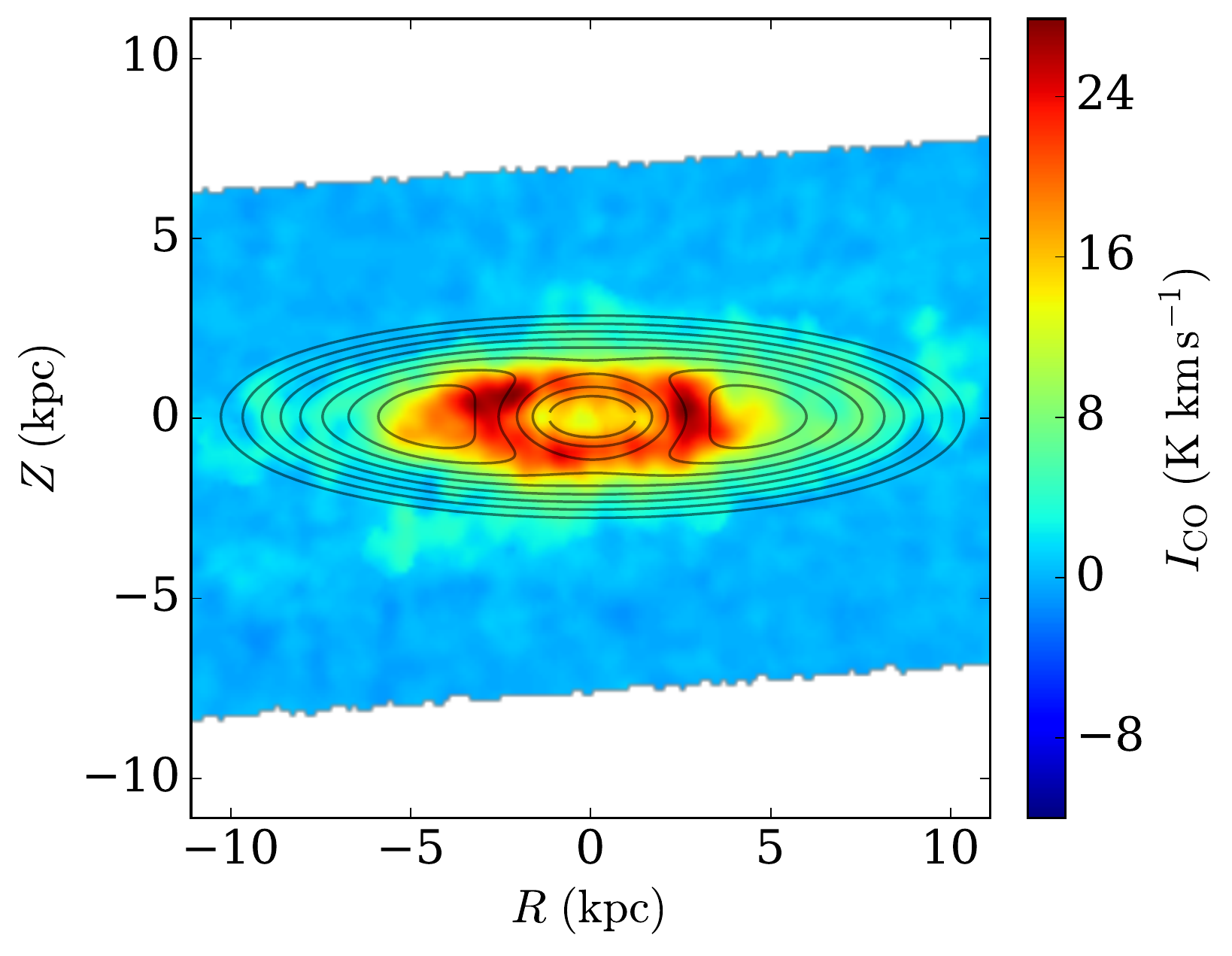}}\\
\end{tabular}
\end{center}
\caption{Top panel: The simulated column density map of the molecular disk. Bottom panel: Overplot of observed (colour scale) and simulated (contour) molecular gas column density maps. The colour bars indicate the observed $\rm I_{CO}$ in the unit of K \kms. The successive contours are for $\rm I_{CO} = $ (8, 10, 12, 14, 16, 18, 20, 22, 24) K \kms.}
\label{mom0_3comp}
\end{figure}

\begin{figure}
\begin{center}
\begin{tabular}{c}
\resizebox{.4\textwidth}{!}{\includegraphics{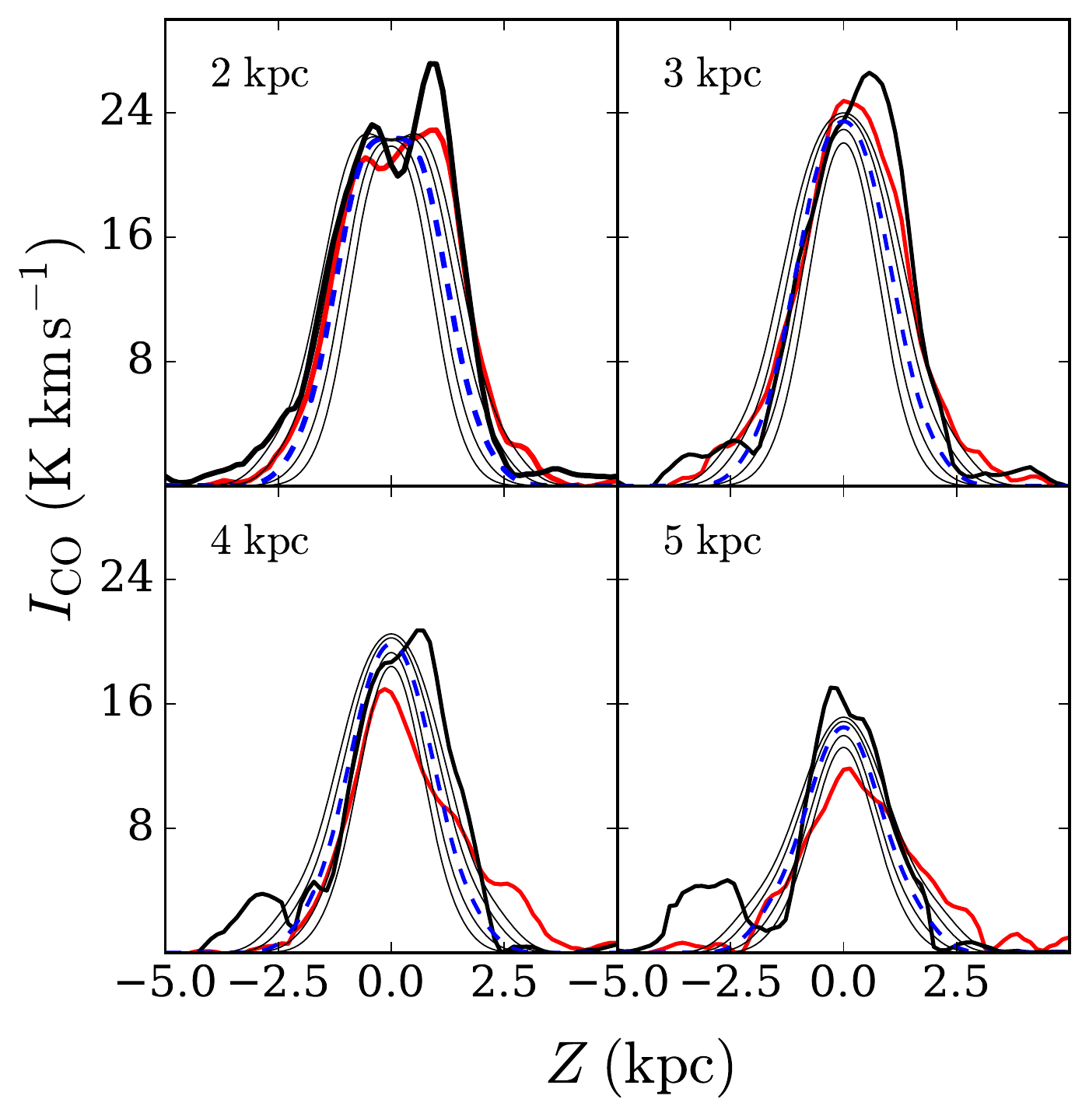}}
\end{tabular}
\end{center}
\caption{The vertical profiles of the observed and simulated molecular column density maps at different radial distances (as mentioned at the top left corner of each panel). The solid red and black curves represent the observed vertical profiles of two different halves at a particular galactocentric radius. The blue dashed line represents the vertical profile of the modelled molecular disk.}
\label{vprof_3comp}
\end{figure}

In Fig.~\ref{mom0_3comp} top panel, the simulated column density map of the molecular gas in NGC 7331 is shown as is obtained using the solutions of Eq.~\ref{eq_hydro}. In the bottom panel, this simulated map is compared with the observed one. A $\sigma_{H_2} = 7$ \kms~was adopted to solve the hydrostatic equation. The simulated and the observed molecular disk found to match very well with each other. However, one should exclude central two kpc region for comparison as the Eq.~\ref{eq_hydro} was not solved at $R \lesssim$ 1 kpc and the convolution with the telescope beam of size $\sim$ 1 kpc will introduce error. A careful inspection of the bottom panel of the figure reveals extra emission features at both the upper and lower half of the galaxy which was not accounted by the model-observation (contours) properly. 
 
To capture it in more details, instead of an overplot of column density maps, I estimate and compare the observed and the modelled vertical profiles of the column density distribution maps as a function of radii. To extract the profiles, a vertical cut along the minor axis was taken through the observed and modelled molecular disk. In Fig.~\ref{vprof_3comp} the vertical profiles are plotted as a function of height ($Z$) from the major axis. It can be seen from the figure that, the observed and the modelled profiles matches with each other reasonably well, though, the base of the observed profiles are somewhat fatter than the modelled one. From Fig.~\ref{mom0_3comp} bottom panel also one can see the extra plume of emissions in the edges of the map surpassing the contours representing the model.

To check if the assumed inclination of the molecular disk is producing this difference, I re-computed the vertical profiles for a set of inclinations. The molecular disk was produced for an inclination range of 72$^o$ to 80$^o$ in steps of 2$^o$. The face-on surface densities of the baryonic disks were recalculated assuming the updated inclination and the hydrostatic equilibrium equation was solved to produce the simulated molecular disks. The thin black curves in each panel of Fig.~\ref{vprof_3comp} represents vertical profiles for inclinations 72$^o$ to 80$^o$ in step of 2$^o$. The outermost profile is for 72$^o$ whereas the innermost profile is for 80$^o$. As can be seen from the figure, no particular inclination can be considered as a better replacement for the assumed inclination of 76$^o$ as they all seem to be consistent given the spatial resolution of $\sim$ 1 kpc. For further analysis, I assume an inclination of 76$^o$ only (observed inclination of the optical disk). It is possible that the extra emission observed at the edges of the molecular disk is produced by a warp at the outer disk as the existence of warps is very common in large galaxies. In such cases, the hydrostatic equilibrium will not work, and the solutions would produce results unmatched to the observation. However, the warps are observed mostly at the outer disks, and it does not perturb the stability of the entire disk.

\begin{figure}
\begin{center}
\begin{tabular}{c}
\resizebox{0.4\textwidth}{!}{\includegraphics{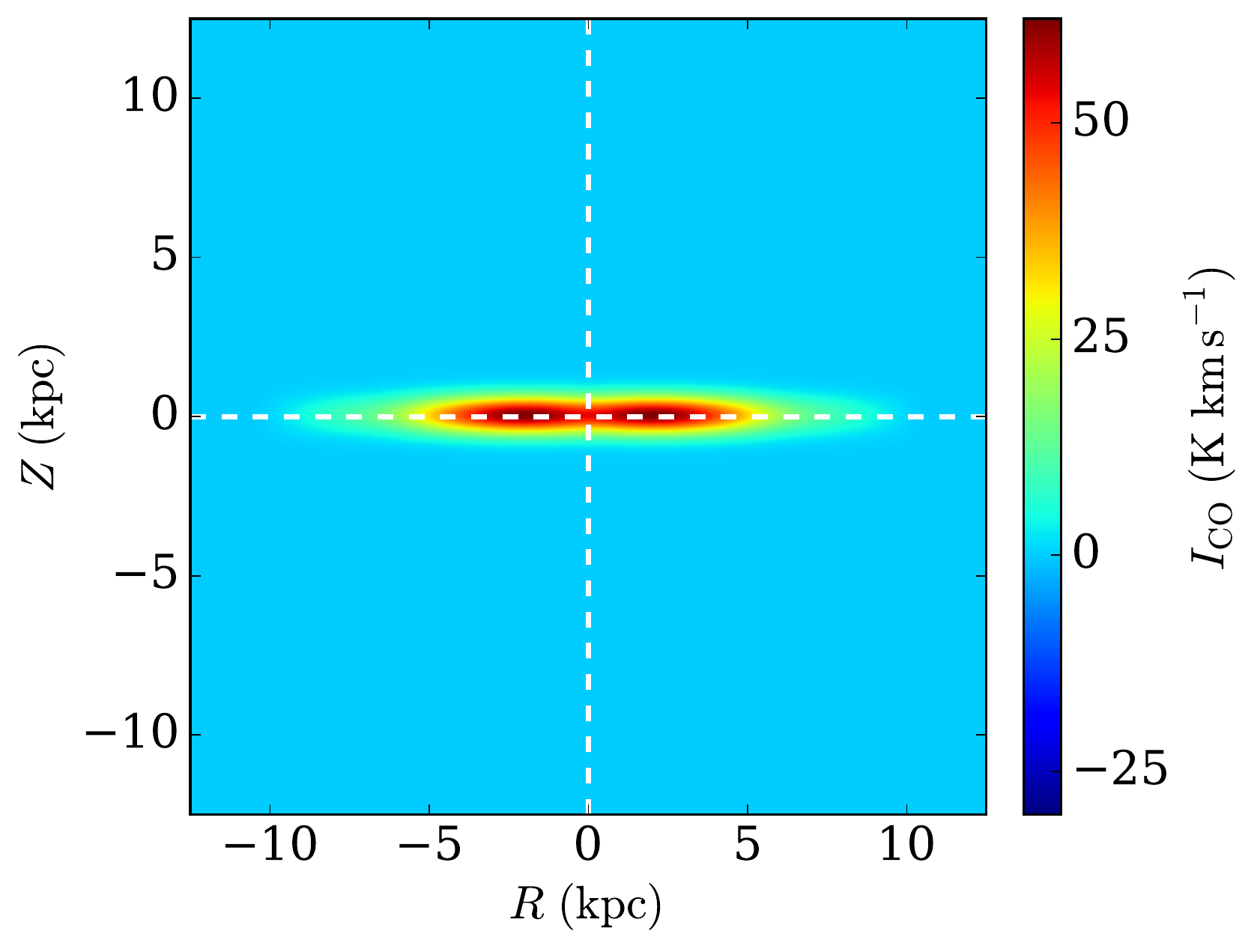}}\\
\resizebox{0.35\textwidth}{!}{\includegraphics{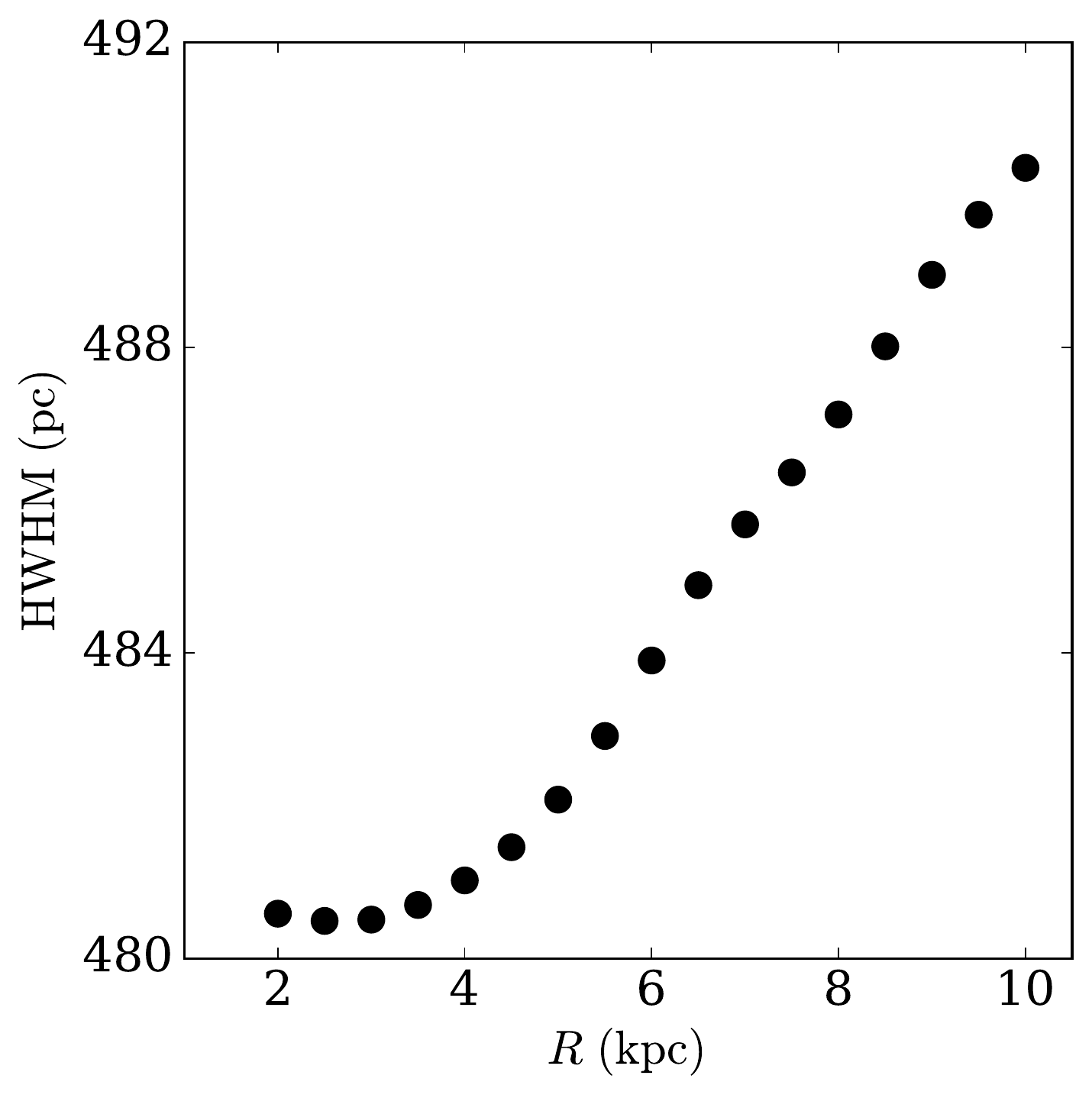}}\\
\end{tabular}
\end{center}
\caption{Top panel: The simulated column density map of molecular disc as seen edge-on. Bottom panel: The HWHM profile of the simulated edge-on column density map. As can be seen the HWHM profile marginally flares as a function of radius and has a value $\sim 500$ pc.}
\label{mom0_incl90}
\end{figure}

As discussed in \S~\ref{intro}, a thick molecular disk extending up to a few kpc from the mid-plane were observed in many edge-on galaxies \citep{garcia-burillo92}. The existence of such thick disks are puzzling, and their origins are still an open question. I extend this study further to check if NGC 7331 in hydrostatic equilibrium can produce an observationally thick molecular disk. To check that, the solved density distribution of the molecular disk of NGC 7331 was projected to an inclination of 90$^o$ to estimate its surface density map at edge-on orientation. In Fig.~\ref{mom0_incl90} top panel, I show the molecular disk of NGC 7331 at an edge-on projection, whereas the bottom panel shows the HWHM profile of the vertical profiles of this disk as a function of radii. With radii, this HWHM profile marginally flares with a value of $\sim 500$ pc. However, this HWHM profile, in fact, can produce a detectable molecular disk of $\sim$ 2 kpc (considering both the sides of the midplane) when observed edge-on. Here it should be noted that the scale height of the molecular gas ($\sim 100-200$ pc) does not represent the full extent of the molecular disk, rather it is an indicative measure of thickness which marks the width where the density falls to half of the maximum. Though the thickness of the molecular disk found in NGC 7331 is not as thick as seen in NGC 891 (which is a few kpc thick~\citep{garcia-burillo92}), these results indicate that it is probably not very difficult to produce a thick molecular disk under the assumption of hydrostatic equilibrium.

\section{Summary and future work}
\label{summary}

In summary, assuming a vertical hydrostatic equilibrium between different disk components in a galaxy, the combined Poisson's-Boltzman equations were set-up and solved to calculate the vertical structure of the molecular disk in the galaxy NGC 7331. Three coupled second-order partial differential equation (Eq.~\ref{eq_hydro}) were solved numerically using 8$^{th}$ order Runge-Kutta method from {\tt scipy} and was implemented using MPI based code for fast computation. For NGC 7331, the hydrostatic equation was solved at $1 \leq R \leq 10$ kpc to obtain the vertical structure of the molecular disk. The molecular scale height was found to be $\sim 50-100$ pc at the centre which increases to $\sim 100-200$ pc at the outer edge. The molecular scale height is sensitive to the assumed $\sigma_{H_2}$ and found to change by a factor of $\sim 2$ when $\sigma_{H_2}$ changes from 6 \kms~to 10 \kms. 

Using the solutions of hydrostatic equations and the observed rotation curve, a three dimensional dynamical model of the molecular disk was made. This model was then inclined to the observed inclination and convolved with the telescope beam to produce a model intensity map. This model intensity map was then compared with the observed one (Fig.~\ref{mom0_3comp}) to find that the model matches with the observation reasonably. However,  some low-intensity excess emission features in the observed molecular map at largest heights was not modelled properly. This emission feature was observed at the edge of the molecular disk and most probably not a part of the stable molecular disk, or it could be a warp. To check the modelling of the molecular disk at different depths, vertical profiles of column density maps were extracted and compared. The modelled vertical profiles at different radii reasonably match with the observation. The effect of assumed inclination on the molecular disk also explored and found that the vertical profile at any radius does not show large change as compared to the observing beam as one changes the inclination from $\sim$ 72$^o$ to 80$^o$.

Finally, I project the molecular gas density distribution of NGC 7331 to an inclination of 90$^o$ to examine if it can produce a reasonably thick molecular disk. The extracted HWHM profile of this edge-on disk was found to be $\sim 500$ pc with a very little flaring with radius. This HWHM was found to be capable of producing a thick observable disk of thickness $\sim$ 2 kpc. With this result, it appears that a simple vertical hydrostatic model of the molecular disk can in-principal produce a few kilo-parsec thick observed disk and hence, creating a thick molecular disk in external galaxies might not be as difficult as it was thought before (e.g., NGC 891 \citep{garcia-burillo92}).

In this work I assumed the molecular disk to be a single component system with a single $\sigma_{H_2}$. However, as discussed in \S~\ref{intro}, many recent studies point towards the possibility of a two-component molecular disk with a thin disk residing close to the midplane and a diffuse thick disk extending up to a few kpc. The $\sigma_{H_2}$ of these disks are expected to be different. In these scenarios, the assumption of a simple single component molecular disk will fail, and one needs to add an extra component to the hydrostatic equilibrium equation. In future work, a detailed study with a two-component molecular disk is worth exploring to understand the thick molecular disks observed in external galaxies.

\section{Acknowledgement}
NNP would like to thank Dr Yogesh Wadadekar, Dr Samir Choudhuri and Mrs Gunjan Verma for their comments and suggestions which helped to improve the quality of this manuscript. NNP would also like to thank both the referees for their valuable comments and suggestions which improved the quality and readability of this paper.

\bibliographystyle{mn2e}
\bibliography{bibliography}

\end{document}